\documentclass[twocolumn,epjc3]{svjour3}

\smartqed
\RequirePackage{amsmath,graphicx}
\RequirePackage{latexsym}
\RequirePackage[colorlinks,citecolor=blue,urlcolor=blue,linkcolor=blue]{hyperref}

\RequirePackage{lineno,isotope,mathrsfs}
\RequirePackage{bbold,amssymb, color}
\RequirePackage{nicefrac,exscale,multirow,times,txfonts}
\RequirePackage{ulem}

\newcommand{\ivec}[1]{\vec{#1}}
\newcommand{\svec}[1]{\boldsymbol{#1}}

\newcommand{\red}[1]{\textcolor[rgb]{1.00,0.00,0.00}{#1}}

\newcommand{\delete}{\bgroup\markoverwith{\textcolor{red}{\rule[0.5ex]{2pt}{1pt}}}\ULon}

\renewcommand{\insert}[1]{\red{#1}}
\newcommand{\ff}[1]{\frac{1}{#1}}

\newcommand{\lrlc}[1]{\left|#1\right>}

\journalname{Eur. Phys. J. A}
\begin{document}

\title{Deformed ground state of $^{32}$Mg and breaking of pseudo-spin symmetry}
%\subtitle{Do you have a subtitle?\\ If so, write it here}

%\titlerunning{Short form of title}        % if too long for running head

\author{Yong Peng \thanksref{addr1,addr2}\and
        Jing Geng \thanksref{e1,addr1,addr2}\and
        Yi Fei Niu \thanksref{addr1,addr2,addr3}\and
        Wen Hui Long \thanksref{e2,addr1,addr2,addr3}
}

%\thankstext{t1}{Grants or other notes
%about the article that should go on the front page should be
%placed here. General acknowledgments should be placed at the end of the article.
\thankstext{e1}{e-mail: gengj17@lzu.edu.cn}
\thankstext{e2}{e-mail: longwh@lzu.edu.cn}
%\authorrunning{Short form of author list} % if too long for running head

\institute{Frontier Science Center for Rare isotope, Lanzhou University, Lanzhou 730000, China \label{addr1}\and
           School of Nuclear Science and Technology, Lanzhou University, Lanzhou 730000, China \label{addr2}\and
           Joint Department for Nuclear Physics, Lanzhou University and Institute of Modern Physics, CAS, Lanzhou 730000, China\label{addr3}
}
\date{Received: date / Accepted: date}
% The correct dates will be entered by the editor

\maketitle
\begin{abstract}
Deformed ground state of $^{32}$Mg is investigated using the axially deformed relativistic Hartree-Fock-Bogoliubov (D-RHFB) model with the effective Lagrangian PKA1, which provides coincident description with the experimental measurements. It is illustrated that obvious breaking of the pseudo-spin symmetry (PSS) given by PKA1, being consistent with the experimental observation in nearby isotone $^{40}$Ca, is crucial for describing correctly the deformed ground state by producing unique shape evolution of neutron orbit $1/2_4^+$ in $^{32}$Mg. The PSS breaking is essentially determined by characteristic in-medium balance between nuclear attractions and repulsions that is manifested as unparalleled density dependent behaviors for coupling strengths $g_\sigma$ and $g_\omega$ in dominant $\sigma$-scalar and $\omega$-vector channels.
%\keywords{First keyword \and Second keyword \and More}
% \PACS{PACS code1 \and PACS code2 \and more}
% \subclass{MSC code1 \and MSC code2 \and more}
\end{abstract}
%
% Computer program descriptions must contain the following
% PROGRAM SUMMARY AND SPECIFICATIONS.
%\noindent
%{\bf Program Summary and Specifications}\\
%%Delete as appropriate.
%\begin{small}
%\noindent
%{Program title:}\\
%{Licensing provisions:}\\
%%Please choose one among the following:  CC0 1.0/CC By 4.0/MIT/Apache-2.0/BSD 3-clause/BSD 2-clause/GPLv3/GPLv2/LGPL/CC BY NC 3.0/MPL-2.0
%{Programming language:}\\
%{Repository and DOI:}\\
%%Please indicate where the program and its additional files have been deposited (e.g. Github,� , together with the DOI obtained.
%{Description of problem:}\\
%%Describe the nature of the problem here in approx. 50-250 words.
%{Method of solution:}\\
%%Describe the method solution here in approx. 50-250 words.
%{Additional comments:}\\
%%Provide any additional comments (e.g. previous versions, new version information and summary, unusual features, limitations,� here in approx. 50-250 words.
%\end{small}
%
\section{Introduction}
Nuclei with extreme neutron-proton ratio, called as exotic nuclei or unstable ones, exhibit rich novel nuclear phenomena. {As one of the typical examples}, the alterations of microscopic structures can give rise to the disappearance of traditional magic shells and the occurrence of new ones, which may accompany with stable nuclear deformation. The nucleus $^{32}$Mg, located at so-called island of inversion \cite{Warburton1990PRC41.1147}, has received wide attentions due to its extraordinary structural properties \cite{Caurier2005RMP77.427,Sorlin2008PPNP61.602}, i.e., vanishing neutron magic shell $N=20$ and stably deformed ground state.

Experimentally, the large $B(E2:0_{\text{g.s.}}^+\to 2_{1}^{+})$ values extracted from Coulomb excitation measurements \cite{Motobayashi1995PLB346.9,Pritychenko1999PLB461.322,Church2005PRC72.054320} establish stably deformed ground state of $^{32}$Mg, being consistent with  measured low-lying $2_1^+$ state \cite{Pritychenko1999PLB461.322,GUILLEMAUD1984PRC426.37,Church2005PRC72.054320}. Coincidentally, distinct quadruple collectivity of $^{32}$Mg was revealed by measured ratio $E(4_1^+)/E(2_1^+)=2.6$ that lies between the vibration limit 2.0 and rigid rotor limit 3.3 \cite{Takeuchi2009PRC79.054319}, and later confirmed by observed rotational band structures of $^{32}$Mg \cite{Crawford2016PRC93.031303(R)}, a characteristic fingerprint of a rigid non-spherical shape. Moreover, from observed population of excited $0_2^{+}$ state from two-neutron transfer reaction $(\text{t, p})$ on $^{30}$Mg \cite{Wimmer2010PRL105.252501}, it was suggested that the excited $0_2^{+}$ state is spherical, coexisting with deformed ground state of $^{32}$Mg.

In parallel with the experimental studies, theoretical efforts were devoted to the quadruple collective dynamics of $^{32}$Mg by applying the shell model \cite{Caurier2005RMP77.427,Utsuno1999PRC60.054315,Caurier2001NPA693.374,Otsuka2004EPJA20.69,Caurier2014PRC90.014302}, {quasi-particle random phase approximation established on the Hartree-Fock-Bogoliubov scheme} \cite{Yamagami2004PRC69.034301}, and generator coordinate method based on the relativistic mean-field (RMF) and non-relativistic mean-field models \cite{Yao2011PRC83.014308,Li2012PRC85.024312,Rodr2002NPA709.201}. {After introducing neutron $2p$-$2h$ excitation across the $sd$-$pf$ shell gap, the shell model calculations can interpret the large $B(E2)$ values and low-lying $2_1^+$ state}. {Similarly, only going beyond mean-field, the low-lying $2_1^+$ state can be reproduced by the non-relativistic or relativistic mean field models.} However, the deformed ground state of $^{32}$Mg was neither supported directly nor described self-consistently by {the mentioned} models. Furthermore, the deformed $N = 20$ shell structure and microscopic evidence for deformed ground state for $^{32}$Mg were not clarified yet. As one of the representative models, the RMF theory \cite{Walecka1974Ann.Phys83.491,Reinhard1989RPP52.439}, that contains only the Hartree diagram of the meson-propagated nuclear force \cite{Yukawa1935Proc.Phys.Math.Soc.Japan17.48}, has achieved great success in describing various nuclear phenomena \cite{Ring1996PPNP37.193,Bender2003RMP75.121,Vretenar2005PRe409.101,Meng2006PRC73.037303,Meng2006PPNP57.470,Niksic2011Prog.Part.Nucl.Phys66.519,Liang2015PRe570.1}. It shall be emphasized that for the RMF approach, appropriate modeling of nuclear in-medium effects, via either the nonlinear self-couplings of mesons \cite{Boguta1977NPA292.413,Sugahara1994NPA579.557,Long2004PRC69.034319} or the density dependencies of the meson-nucleon coupling strengths \cite{Brockmann1992PRL68.3408,Lenske1995PLB345.355,Fuchs1995PRC52.3043,Typel1999NPA656.331,Long2004PRC69.034319}, is essential for providing accurate and reliable descriptions, e.g., improved saturation properties of nuclear matter.

Implementing the Fock diagram of the meson-propagated nuclear force, the relativistic Hartree-Fock (RHF) descriptions \cite{Bouyssy1987PRC36.380} of nuclear structure were also improved by modeling the nuclear in-medium effects via the non-linear self-couplings of $\sigma$-meson \cite{Bernardos1993PRC48.2665} and scalar field $\big(\bar\psi\psi\big)$ \cite{Marcos2004JPG30.703}. Eventually, similar accuracy as popular RMF models was achieved by the density-dependent relativistic Hartree-Fock (DDRHF) theory \cite{Long2006PLB640.150,Long2007PRC76.034314} and relativistic Hartree-Fock-Bogoliubov (RHFB) theory \cite{Long2010PRC81.024308}, {in which the meson-nucleon coupling strengths are assumed to be density dependent for modeling the nuclear in-medium effects}. Moreover, significant improvements due to the Fock terms were continuously found in the self-consistent descriptions of shell evolution \cite{Long2008EPL82.12001,Long2009PLB680.428,Wang2013PRC87.047301}, new magicity \cite{Li2014PLB732.169,Li2016PLB753.97,Li2019PLB788.192,Liu2020PLB806.135524}, the effects of tensor force \cite{Jiang2015PRC91.025802,Jiang2015PRC91.034326,Zong2018CPC42.024101,Wang2020PRC101.064306,Geng2020PRC101.064302}, nuclear spin-isospin excitations \cite{Liang2008PRL101.122502,Liang2012PRC85.064302,Wang2020PRC101.064306}, etc.

Benefited from the covariant form with attractive scalar potential $S(r)$ and repulsive vector one $V(r)$, the RMF models, as well as the RHF ones, work well on describing not only the strong spin-orbit coupling, but also the origin of the pseudo-spin symmetry (PSS) \cite{Hecht1969NPA137.129,Arima1969PLB30.517,Ginocchio1999PhyRep315.231,Long2006PLB639.242,Liang2015PRe570.1}, quasi-degeneracy of the so-called pseudo-spin (PS) doublet ($n$, $l$, $j = l + 1/2$) and ($n-1$, $l + 2$, $j = l + 3/2$). Under the RMF scheme, the conservation condition of the PSS was demonstrated as $S(r)+V(r)=0$ \cite{Ginocchio1997PRL78.436} or $d[S(r) + V(r)]/dr=0$ \cite{Meng1998PLB419.1}, both of which indeed indicate certain in-medium balance between nuclear attractions and repulsions. Regarding the covariant representation of the RMF and RHF models, such in-medium balance is determined mainly by attractive $\sigma$-scalar ($\sigma$-S) and repulsive $\omega$-vector ($\omega$-V) couplings, {and more specifically it can} be {shaped by} the density dependencies of the coupling strengths{, showing a tight relation to the modeling of} the nuclear in-medium effects. {For instance}, significantly improved in-medium balance has been achieved by the RHF Lagrangian PKA1 with strong $\rho$-tensor ($\rho$-T) coupling, showing as unparalleled density dependent behaviors for the coupling strengths $g_\sigma$ and $g_\omega$ respectively in the $\sigma$-S and $\omega$-V channels \cite{Long2007PRC76.034314}, in contrast to popular RMF models and RHF ones with PKO$i$ ($i=1,2,3$). Recently, it was illustrated in Ref. \cite{Geng2019PRC100.051301R} that improved in-medium balance is crucial to reproduce the PSS restoration of the high-$l$ PS doublets \cite{Long2007PRC76.034314,Long2009PLB680.428,Long2010PRC81.031302,Wei2020CPC44.074107}.

More recently, utilizing the spherical Dirac Woods-Saxon (DWS) base \cite{Zhou2003PRC68.034323}, both RHF and RHFB models were extended for axially deformed nuclei \cite{Geng2020PRC101.064302,Geng2022PRC105.034329}, leading to the D-RHF and D-RHFB models, respectively. It inspires us to verify the underlying mechanism behind the unusual stably deformed ground state of $^{32}$Mg, also regarding the full advantages achieved by the D-RHFB model on unified treatments of the spin-orbit coupling, tensor force, deformation, pairing correlations and continuum effects \cite{Geng2022PRC105.034329}. The paper is organized as below, in Sec. \ref{sec:General Formalism}, the general formalism is briefly recalled. Afterwards, the relation between deformed ground state of $^{32}$Mg and breaking of pseudo-spin symmetry are discussed in Sec. \ref{sec:Results}. Finally, a summary is given in Sec. \ref{sec:Summary}.

\section{General Formalism}\label{sec:General Formalism}

Restricted with the mean field approach, both RMF and RHF models are established on the meson-propagated picture of nuclear force.
Specifically, the isoscalar $\sigma$-S and $\omega$-V couplings dominate nuclear attractions and repulsions, respectively, and the isovector $\rho$-vector ($\rho$-V), $\rho$-T, $\rho$-vector-tensor ($\rho$-VT) and  $\pi$-pseudo-vector ($\pi$-PV) couplings account for the isospin-related properties of nuclear force and partly the tensor force effects, and the photon-vector ($A$-V) coupling for the Coulomb repulsions between protons \cite{Bouyssy1987PRC36.380,Long2006PLB640.150,Long2007PRC76.034314,Long2010PRC81.024308,Geng2020PRC101.064302,Geng2022PRC105.034329}. {Thus, the Lagrangian that describes nuclear systems can be deduced as the theoretical starting point, from which the Legendre transformation gives the RHF Hamiltonian as},
\begin{equation}\label{eq:Hamiltonian}
    H = T + \sum_\phi  V_\phi,
\end{equation}
{with the kinetic energy term ($T$) and potential energy ones ($ V_\phi$) reading as},
\begin{align}
     T =& \int d\svec x \bar\psi(\svec x) \left(-i\svec\gamma \cdot\svec \nabla + M\right) \psi(\svec x), \label{eq:kinetic-O}\\
     V_\phi = & \ff2 \int d\svec x d\svec x' \bar\psi(\svec x) \bar\psi(\svec x') \Gamma_\phi D_\phi(\svec x-\svec x') \psi(\svec x') \psi(\svec x).\label{eq:potential-O}
\end{align}
{In the above expressions,} $\psi$ {represents} Dirac spinor field, $\phi$ {denotes} various two-body interaction channels, namely the $\sigma$-S, $\omega$-V, $\rho$-V, $\rho$-T, $\rho$-VT, $\pi$-PV and $A$-V couplings{, and the interaction vertex $\Gamma_\phi( x, x')$ read as,}
\begin{subequations}\label{eq:vertex}
\begin{align}
    \Gamma_{\sigma\text{-S}} \equiv & -g_\sigma(x) g_\sigma(x'), \\
    \Gamma_{\omega\text{-V}} \equiv & \left(g_\omega \gamma_\mu\right)_{x} \left( g_\omega\gamma^\mu \right)_{x'}, \\
    \Gamma_{\rho\text{-V}} \equiv & \left( g_\rho \gamma_\mu \ivec\tau \right)_x \cdot \left( g_\rho \gamma^\mu \ivec\tau \right)_{x'}, \\
    \Gamma_{\rho\text{-T}} \equiv & \frac{1}{4M^2} \left( f_\rho \sigma_{\nu k} \vec\tau \partial^k \right)_{x} \cdot \left( f_\rho \sigma^{\nu l} \vec\tau \partial_l\right)_{x'}, \\
    \Gamma_{\rho\text{-VT}} \equiv & \frac{1}{2M} \left( f_\rho \sigma^{k\nu} \vec\tau \partial_k \right)_{x} \cdot \left( g_\rho \gamma_\nu \vec\tau \right)_{x'} \nonumber \\
    & + \frac{1}{2M}  \left( g_\rho \gamma_\nu \vec\tau \right)_{x} \cdot \left( f_\rho \sigma^{k\nu} \vec\tau \partial_k\right)_{x'}, \\
    \Gamma_{\pi\text{-PV}} \equiv & \frac{-1}{m_\pi^2} \left( f_\pi \vec\tau \gamma_5 \gamma_\mu \partial^\mu \right)_{x} \cdot \left( f_\pi \vec\tau \gamma_5 \gamma_\nu \partial^\nu \right)_{x'}, \\
    \Gamma_{A\text{-V}} \equiv & \frac{e^2}{4} \left( \gamma_\mu (1- \tau) \right)_{x} \left( \gamma^\mu (1- \tau) \right)_{x'},
\end{align}
\end{subequations}
{where $x = (t, \svec r)$, and $\ivec\tau$ is the isospin operator, $\tau$ for the projection with the conventions $\tau\lrlc{n} = \lrlc{n}$ and $\tau\lrlc{p} = -\lrlc{p}$.} {After neglecting the retardation effects, the propagators $D_\phi(\svec x-\svec x')$ of the meson and photon field in $V_\phi$ can be written as,}
\begin{equation}\label{eq:Yukawa}
    D_\phi = \frac{1}{4\pi} \frac{e^{-m_\phi\left| \svec x-\svec x' \right|}}{\left|\svec x-\svec x'\right|}, \hspace{2em} D_{A} = \frac{1}{4\pi} \frac{1}{\left|\svec x-\svec x'\right|},
\end{equation}
{where $m_\phi$ is the meson mass in the meson-nucleon coupling channel $\phi$.}

{Restricted with the mean field approach, the modeling of nuclear in-medium effects is necessitated for accurate description of nuclear properties. In the utilized D-RHFB model, nuclear in-medium effects are evaluated by introducing the density dependencies into the meson-nucleon coupling strengths, namely $g_\sigma$, $g_\omega$, $g_\rho$, $f_\rho$ and $f_\pi$ in the interaction vertex (\ref{eq:vertex}). Together with the meson masses $m_\sigma$, $m_\omega$, $m_\rho$ and $m_\pi$, the coupling strengths and their density dependencies define an in-medium interaction for nuclear systems.}

{For the isoscalar $\sigma$-S and $\omega$-V coupling strengths, the density dependencies read as,}
\begin{align}
  g_\phi = & g_\phi(\rho_0) f_\phi(\xi), & f_\phi(\xi) = & a_\phi\frac{1+b_\phi(\xi+d_\phi)^2}{1+c_\phi(\xi+d_\phi)^2},
\end{align}
{where $\xi = \rho_b/\rho_0$, $\rho_0$ being the saturation density and $\rho_b=\bar\psi\gamma_0\psi$ for nucleon density, $\phi$ represents $\sigma$-S and $\omega$-V couplings, and $a_\phi$, $b_\phi$, $c_\phi$ and $d_\phi$ define the density dependencies of the coupling strengths.} {For the isosvector channels, the density dependencies are of the following exponential form,}
\begin{align}
  g_\rho = & g_\rho(\rho_0) e^{-a_\rho(\xi -1)}, & f_{\phi'} = & f_{\phi'}(\rho_0) e^{-a_{\phi'}(\xi -1)},
\end{align}
{where $\phi'$ represents the $\pi$-PV and $\rho$-T coupling strengths, and $a_\rho$, $a_\pi$ and $a_T$ describe the density dependencies. }

{In general, nuclear energy functional, whose variation gives equations of motion of nucleons, corresponds to the expectation of RHF Hamiltonian (\ref{eq:Hamiltonian}) with respect to a nuclear many-body state. Thus, one needs to quantize the Dirac spinor field $\psi$ in a carefully chosen space, which is essential for defining the nuclear many-body state as well. Under the RHF approach,} the Dirac spinor field $\psi$ can be quantized as,
\begin{align}\label{eq:expansionHF}
  \psi(x) = & \sum_{l} \psi_l(x)  c_l, &
  \bar\psi(x) = & \sum_{l} \bar\psi_{l}(x)  c_l^\dag,
\end{align}
where the creation and annihilation operators $c_l^\dag$ and $c_l$ {in the Hartree-Fock (HF) space are defined by the solutions of Dirac equation, and $\psi_l(x)$ denotes single-particle (s.p.) wave function with the index $l$ specified for the s.p. states in this paper. Being consistent with the RHF approach, only positive energy states are considered for the quantization (\ref{eq:expansionHF}), leading to so-called no-sea approximation. As a result, the nuclear many-body state under the RHF approach, namely the HF ground state, can be deduced as,}
\begin{align}
  \lrlc{\text{HF}} = & \prod_{l=1}^A c_l^\dag \lrlc{-}, & c_l \lrlc{-} = & 0,
\end{align}
{where $\lrlc{-}$ represents the vacuum state, and $A$ is nuclear mass number. With respect to $\lrlc{\text{HF}}$, the expectation of the Hamiltonian (\ref{eq:Hamiltonian}) gives RHF energy functional \cite{Bouyssy1987PRC36.380,Geng2020PRC101.064302}.}

{When extending from stable to unstable nuclei, the low-lying continuum states can be gradually involved by the pairing correlations, and unified treatment of the RHF mean field and pairing correlations becomes necessitated for the reliable description, which can be achieved within the Bogoliubov scheme with improved self-consistence. Following the Bogoliubov transformation from the HF s.p. space to the Bogoliubov quasi-particle one, }
\begin{equation}\label{eq:Bogoliubov}
    \begin{pmatrix} \beta_k \\[0.5em] \beta_k^\dag \end{pmatrix} = \sum_l \begin{pmatrix} U_{lk}^* & V_{lk}^* \\[0.5em] V_{lk} & U_{lk} \end{pmatrix} \begin{pmatrix} c_l \\[0.5em] c_l^\dag \end{pmatrix},
\end{equation}
with $\beta_k$ and $\beta_k^\dag$ being respectively the annihilation and creation operators {of quasi-particles, and further combining with the quantization (\ref{eq:expansionHF}), a new quantization of Dirac spinor field $\psi(x)$ was proposed in Ref. \cite{Geng2022PRC105.034329} as},
\begin{subequations}\label{eq:expansionHFB}
\begin{align}
  \psi(x) = & \sum_k \big(\psi_{\bar k}^{U}(\svec r) e^{-i\varepsilon_k t} \beta_k + \psi_k^V(\svec r) e^{+i\varepsilon_k t} \beta_k^\dag\big), \\
  \bar\psi(x) = & \sum_k \big(\bar\psi_k^V(\svec r) e^{-i\varepsilon_k t} \beta_k + \bar\psi_{\bar k}^{U}(\svec r) e^{+i\varepsilon_k t}\beta_k^\dag\big),
\end{align}
\end{subequations}
where $\varepsilon_k$ is quasi-particle energy{, $\psi^V$ and $\psi^U$ are the $V$- and $U$-components of quasi-particle spinors, and the index $k$ and $\bar k$ correspond to the state and its time-reversal partners to form a Cooper pair.}

%\begin{widetext}
{With newly proposed quantization form (\ref{eq:expansionHFB}), the kinetic and potential energies terms, i.e., the $T$ and  $V_\phi$ in the Hamiltonian (\ref{eq:Hamiltonian}) can be expressed as},
\begin{subequations}\label{eq:Hamiltonian-HFB}
\begin{align}
  T = & \sum_{kk'}\int d\svec r \bar\psi_k^V(\svec r)(-i\svec\gamma\cdot\svec\nabla + M)\psi_{k'}^V (\svec r) \beta_k \beta_{k'}^\dag, \label{eq:Ham-k}\\
  V_\phi = & \ff2 \sum_{k_1k_2 k_2'k_1'} \int d\svec r d\svec r'  \Big[\bar\psi_{k_1}^V(\svec r)
  \bar\psi_{k_2}^V(\svec r')\Gamma_\phi (\svec r, \svec r') D_{\phi}(\svec r-\svec r') \nonumber\\
  &\hspace{3em}\times\psi_{k_2'}^V(\svec r')  \psi_{k_1'}^V(\svec r) \beta_{k_1} \beta_{k_2} \beta_{k_2'}^\dag \beta_{k_1'}^\dag  \nonumber\\
  &\hspace{5em}+ \bar\psi_{k_1}^V(\svec r)\bar\psi_{\bar k_2}^U(\svec r')\Gamma_\phi(\svec r, \svec r') D_{\phi}(\svec r-\svec r')\nonumber\\
  &\hspace{7.5em}\times\psi_{\bar k_2'}^U(\svec r')\psi_{k_1'}^V(\svec r) \beta_{k_1} \beta_{k_2}^\dag \beta_{k_2'}\beta_{k_1'}^\dag\Big]\label{eq:Ham-v}.
\end{align}
\end{subequations}
Obviously, the first term in $V_\phi$ corresponds to the contribution of the mean field, and the second one accounts for the pairing correlations.
%\end{widetext}

{To derive the RHFB energy functional, the Bogoliubov ground state $\lrlc{\text{HFB}}$, that fulfills the condition $\beta_k \lrlc{\text{HFB}}=0$, is taken as the nuclear many-body state. Referring to $\lrlc{\text{HFB}}$, the expectation of the Hamiltonian (\ref{eq:Hamiltonian-HFB}), in which the terms having zero expectations are omitted, leads to a full energy functional containing the kinetic energy $E_{\text{kin.}}$, the potential energy $E_{\text{pot.}}$ and pairing energy $E_{\text{pair}}$ as},
\begin{equation}\label{eq:energy}
  E =  E_{\text{kin.}} + E_{\text{pot.}} + E_{\text{pair}},
\end{equation}
where {the $E_{\text{pot.}}$ term includes the Hartree and Fock contributions} from the $\sigma$-S, $\omega$-V, $\rho$-V, $\rho$-T, $\rho$-VT, $\pi$-PV and $A$-V coupling channels{, and the detailed expressions are referred to Ref. \cite{Geng2022PRC105.034329}. With the obtained energy functional, it is convenient to derive the RHFB equations by performing the variation with respect to the generalized density matrix \cite{Geng2022PRC105.034329}. In general, the pairing energy $E_{\text{pair}}$ is evaluated in a phenomenological way. Here the finite range Gogny force D1S \cite{Berger1984NPA428.23} is adopted as the pairing force, and the $\Gamma_\phi D_\phi$ in the second term of Eq. (\ref{eq:Ham-v}) is replaced by $(\gamma_0)_x(\gamma_0)_{x'} V_{\text{Gogny}}^{pp}(\svec r-\svec r')$ \cite{Geng2022PRC105.034329}.}

{In this work, the studied nucleus $^{32}$Mg is deformed, and the axial symmetry and the reflection one with respect to $z=0$ plane are imposed. Thus, the projection $m$ of} the s.p./quasi-particle angular momentum $j$ and parity $\pi$ remain as good quantum {numbers. For abbreviated expressions,} the index $i=(\nu\pi m)$ is used to denote deformed {s.p./quasi-particle orbits in the following context}, with $\nu$ for the index of the orbits in the $\pi m$-block. {Because of the non-local Fock terms and finite range pairing force, one obtains the integro-differential RHFB equations which are hard to be solved directly in coordinate space. Aiming at the reliable description of unstable nuclei, the quasi-particle spinors $\psi^U$ and $\psi^V$ are then expanded on the spherical DWS base \cite{Zhou2003PRC68.034323,Long2010PRC81.024308,Geng2022PRC105.034329},}
\begin{align}\label{equ:C}
    \psi_{\nu\pi m}^U =& \sum_{n\kappa} C_{n\kappa,i}^U \psi_{n\kappa m}, &
    \psi_{\nu\pi m}^V =& \sum_{n\kappa}C_{n\kappa,i}^V \psi_{n\kappa m},
\end{align}
where the expansion coefficients $C_{n\kappa,i}^U$ and $C_{n\kappa,i}^V$ are restricted as real numbers, {the index $a = (n\kappa)$, together with the projection $m$, denotes} the states in the spherical DWS base {in the following context}, with $\kappa = \pm (j+1/2)$ and $j=l\mp1/2$, $l$ for orbital angular momentum and $n$ for the principle number.

{In terms of the spherical DWS base, the RHFB equations can be derived as,}
\begin{equation}\label{eq:HFB-E}
    \sum_{a'} \begin{pmatrix} -h_{aa'}^{i} +\lambda & \Delta_{aa'}^i \\[0.5em] \Delta_{aa'}^{i} & h_{aa'}^{i}-\lambda \end{pmatrix} \begin{pmatrix} C_{a',i}^U \\[0.5em] C_{a',i}^V \end{pmatrix} = \varepsilon_i \begin{pmatrix} C_{a,i}^U \\[0.5em] C_{a,i}^V \end{pmatrix},
\end{equation}
{where $\varepsilon_i$ is the quasi-particle energy, the chemical potential $\lambda$ is introduced to preserve the particle number on the average, and $h_{aa^{'}}^{i}$ and $\Delta_{aa'}^i$ are respectively the Dirac s.p. Hamiltonian and pairing potential \cite{Geng2022PRC105.034329}.}

{Practically, it is more convenient to perform physical analysis in canonical s.p. space, and the transformation from the quasi-particle to canonical s.p. space can be obtained by diagonalizing the density matrix.} According to {the expansion (\ref{equ:C}) of} $\psi_{\nu\pi m}^V$, the density matrix elements {for the $\pi m$-block} can be expressed as,
\begin{equation}
  \rho_{aa'}^{\pi m} =  \sum_{\nu} C_{a, \nu\pi m}^V C_{a', \nu\pi m}^V.
\end{equation}
{By} diagonalizing the density matrix $\big(\rho_{aa'}^{\pi m}\big)$, the {obtained} eigenvalues correspond to the occupation probabilities $v_i^2$ of canonical s.p. orbits $m_\nu^\pi$, {and the eigenvectors $\widehat D_{\nu}^{\pi m}$, the set of expansion coefficients $D_{n\kappa, i}$ upon the spherical DWS base, define the canonical wave functions $\psi_{\nu\pi m}$ and s.p. energy $E_i$ as},
\begin{align}\label{eq:Eig}
  \psi_{\nu\pi m} =&  \sum_{n\kappa} D_{n\kappa, i} \psi_{n\kappa m}, &   E_i = & \sum_{aa'} D_{a,i} h_{aa'}^{\pi m}D_{a',i}.
\end{align}
{Following variational principle, the relation between the sum of canonical s.p. energies $E_{\text{s.p.}}$, kinetic energy $E_{\text{kin.}}$, potential energy $E_{\text{pot.}}$ and rearrangement term $E_R$ can be derived as},
\begin{equation}\label{eq:sp-energy}
  E_{\text{s.p.}} = \sum_i v_i^2E_i = E_{\text{kin.}} + 2 E_{\text{pot.}} + 2E_R,
\end{equation}
where the $E_R$ term describes the rearrangement effects due to the density dependencies of the
meson-nucleon coupling strengths \cite{Typel1999NPA656.331},
\begin{align}
  E_R = & \frac{1}{2} \int d\svec r \Sigma_R(\svec r) \rho_b(\svec r).
\end{align}
{In order to understand the s.p. role in determining the binding of a nucleus, the RHFB energy functional (\ref{eq:energy}) can be rewritten as,}
\begin{equation}\label{eq:energy1}
  E =   E_{\text{s.p.}} + E_{\text{Re.}} + E_{\text{Oth.}},
\end{equation}
where {the rearrangement term} $E_{\text{Re.}}\equiv -2E_{R}$ and \text{the others} $E_{\text{Oth.}} \equiv - E_{\text{pot.}}+ E_{\text{pair}}$. {Excluding the rest masses of nucleons, one may obtain the nuclear binding energy as $E_B = E-AM$. As usual, the center-of-mass correction $E_{\text{c.m.}}$, that is not involved in the variation procedure, is considered to provide precise $E_B$ values, namely $E_B = E-AM + E_{\text{c.m.}}$, and the other term thus reads as $E_{\text{Oth.}} = - E_{\text{pot.}}+ E_{\text{pair}} + E_{\text{c.m.}}$.}

\section{Results and Discussions}\label{sec:Results}

{In the D-RHFB calculations, the space truncations have to be determined carefully, including} the maximum projection $m_{\max}$, the expansion terms $\lambda_p$ of the density-dependent coupling strengths, see Eq. (46) in Ref. \cite{Geng2020PRC101.064302}, and the configuration space of the spherical DWS base, namely the $n$ and $\kappa$ quantities in Eqs. (\ref{equ:C}). {For $^{32}$Mg,} the $m_{\max}$ values are adopted as $11/2$ and $13/2$ {respectively for even and odd} parity states. {In expanding the coupling strengths, it is accurate enough to consider five expansion} terms $\lambda_p=0,2,4,6,8$. {For the $\kappa$-quantities in the spherical DWS base, the cutoff are decided} as $19/2$ and $21/2$ {respectively} for {even and odd} parity states. {For the $n$-values, the truncations correspond to the energy cutoff $E_\pm^C \pm M$, the positive ($+$) and negative ($-$) ones in the spherical DWS base. After testing calculations, it was determined as $E_+^C=+350.0$ MeV and $E_-^C=-100.0$ MeV, see Refs. \cite{Geng2020PRC101.064302,Geng2022PRC105.034329} for more details}.

{In order to understand the underlying mechanism of deformed ground state of $^{32}$Mg, the RHF Lagrangians PKA1 \cite{Long2007PRC76.034314}, PKO2 \cite{Long2008EPL82.12001} and PKO3 \cite{Long2008EPL82.12001}, and the RMF one DD-ME2 \cite{Lalazissis2005PRC71.024312} are utilized in this work. Specifically, DD-ME2 and PKO2 share the same meson degrees of freedom, including the $\sigma$-S, $\omega$-V, $\rho$-V and $A$-V couplings. On top of that, the $\pi$-PV coupling, that contributes only via the Fock terms, is taken into account by both PKO3 and PKA1, and additionally PKA1 contains the $\rho$-T and $\rho$-VT couplings which play the role mainly via the Fock terms. Besides, it shall be stressed that for all selected models the nuclear in-medium effects are evaluated by the density dependencies of the meson-nucleon coupling strengths. In the pairing channel, the finite-range Gogny force D1S \cite{Berger1984NPA428.23} is adopted as the pairing force in all the calculations.}

\subsection{Deformed ground state of $^{32}\text{Mg}$ and nuclear in-medium effects}

\begin{figure}[h!]\centering
\includegraphics[width=0.92\linewidth]{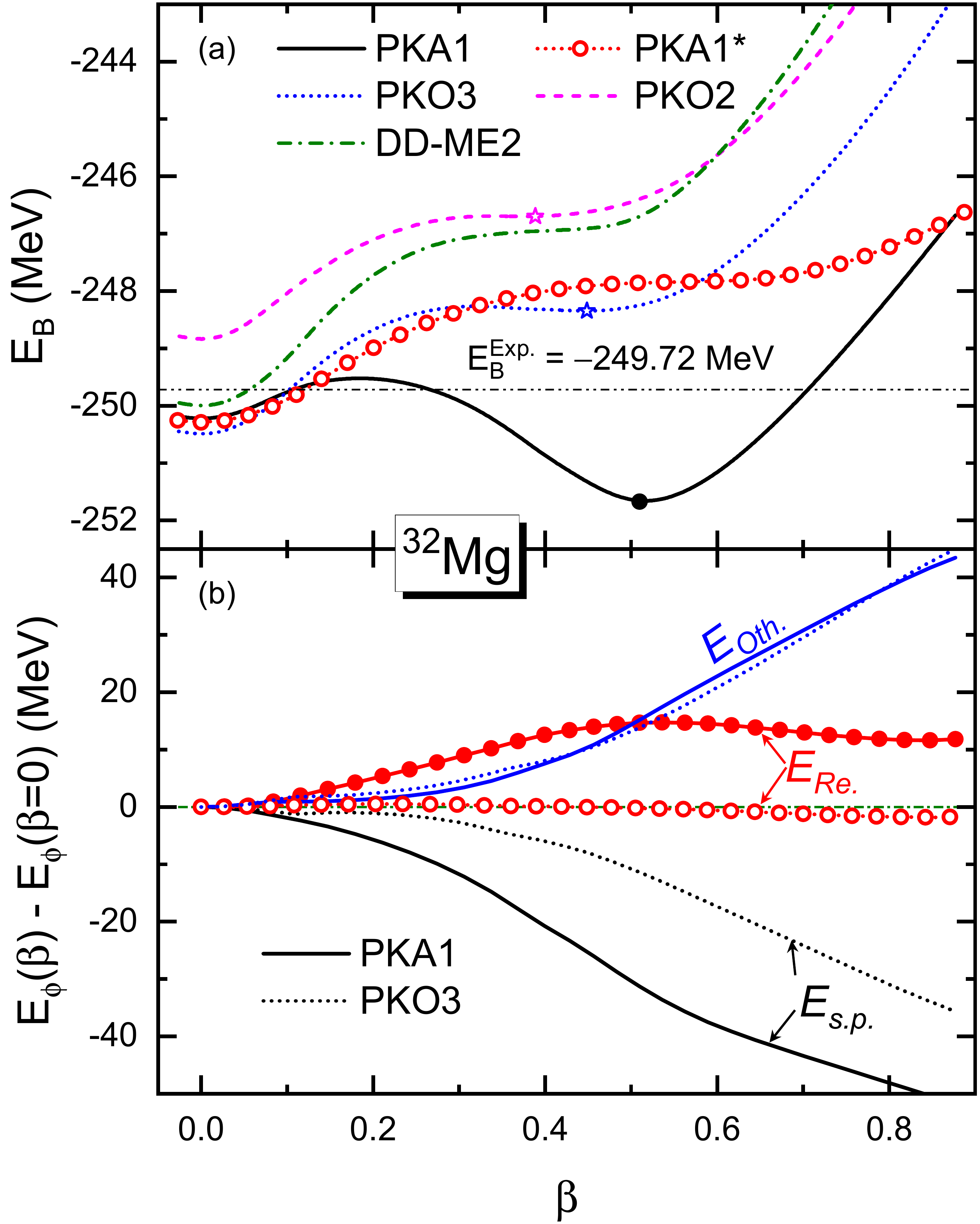}
\caption{(Color Online) Binding energies $E_B$ (MeV) for $^{32}$Mg as functions of quadruple deformation $\beta$ calculated by PKA1, PKO3, PKO2, DD-ME2 and the temporary one PKA1$^{\ast}$. Experimental data is taken form \cite{Wang2017CPC41.030003}. (b) Contributions to the binding energy $E_B$ (MeV) of $^{32}$Mg as functions of quadruple deformation $\beta$ calculated by PKA1 and PKO3, including sum of single particle energy $E_{s.p.}$, the term $E_{Re.}$ and the others $E_{\text{Oth.}}$, in which the values at $\beta$ = 0 are taken as the reference points.}\label{Fig:Ebeta}
\end{figure}

Figure \ref{Fig:Ebeta}(a) shows the binding energy $E_B$ (MeV) of $^{32}$Mg with respect to the quadruple deformation $\beta$, which are extracted from the shape constrained D-RHFB calculations with PKA1, PKO2, PKO3 and DD-ME2. It can be seen that only PKA1 presents the ground state with evident prolate deformation $\beta\approxeq 0.51$ for $^{32}$Mg, {consistent} with the experimental values $\beta=0.512(44)$ \cite{Motobayashi1995PLB346.9} and $\beta=0.51(3)$ \cite{Church2005PRC72.054320}. However, the other selected models give spherical ground state, and  terraces or weak local minima at fairly large prolate deformation. Notice that the possibility of spherical ground state for $^{32}$Mg has been ruled out experimentally \cite{Motobayashi1995PLB346.9,Pritychenko1999PLB461.322,Church2005PRC72.054320,GUILLEMAUD1984PRC426.37,Takeuchi2009PRC79.054319,Wimmer2010PRL105.252501,Crawford2016PRC93.031303(R)}.

{To understand the mechanism behind the deformed ground state, Fig. \ref{Fig:Ebeta} (b) compares the binding energy contributions given by PKA1 and PKO3, including the $E_{\text{s.p.}}$, $E_{\text{Re.}}$ and $E_{\text{Oth.}}$ terms in Eq. (\ref{eq:energy1})}, and the values at $\beta = 0$ are taken as the references. {Under the D-RHFB frame that indicates identical approach on nuclear many-body state for $^{32}$Mg, the deviations between the PKA1 and PKO3 results can only originate from the modeling of nuclear force, such as the considered meson degrees of freedom and the evaluation of nuclear in-medium effects. It is found in Fig. \ref{Fig:Ebeta} (b) that the $E_{\text{Oth.}}$ terms given by PKA1 and PKO3 show rather similar evolution behaviors, in contrast to the $E_{\text{Re.}}$ and $E_{\text{s.p.}}$ ones. Combined with Fig. \ref{Fig:Ebeta} (a), it may indicate that the specific structure of $^{32}$Mg given by PKA1 and PKO3 can be notably different, regarding the fact that the potential energy dominates the $E_{\text{Oth.}}$ terms [see Eq. (\ref{eq:energy1})] and PKA1 contains more coupling channels than PKO3, i.e., the $\rho$-T and $\rho$-VT ones. Moreover, PKA1 and PKO3 provide distinctly different modeling of nuclear in-medium effects \cite{Long2007PRC76.034314,Geng2019PRC100.051301R}. Being consistent with these facts, there exist remarkable difference on the $E_{\text{Re.}}$ and $E_{\text{s.p.}}$ terms given by PKA1 and PKO3 in Fig. \ref{Fig:Ebeta} (b)}. {Specifically, the rearrangement contributions $E_{\text{Re.}}$ given by PKO3 keep near constant, but the ones given by PKA1 show distinct shape dependence when approaching the ground state. Coincidentally, as compared to PKO3, PKA1 presents much more negatively enhanced $E_{\text{s.p.}}$ values following the deformation $\beta$, which seems crucial for PKA1 to describe correctly the deformed ground state of $^{32}$Mg. }

\begin{figure}[h!]\centering
 \includegraphics[width=0.92\linewidth]{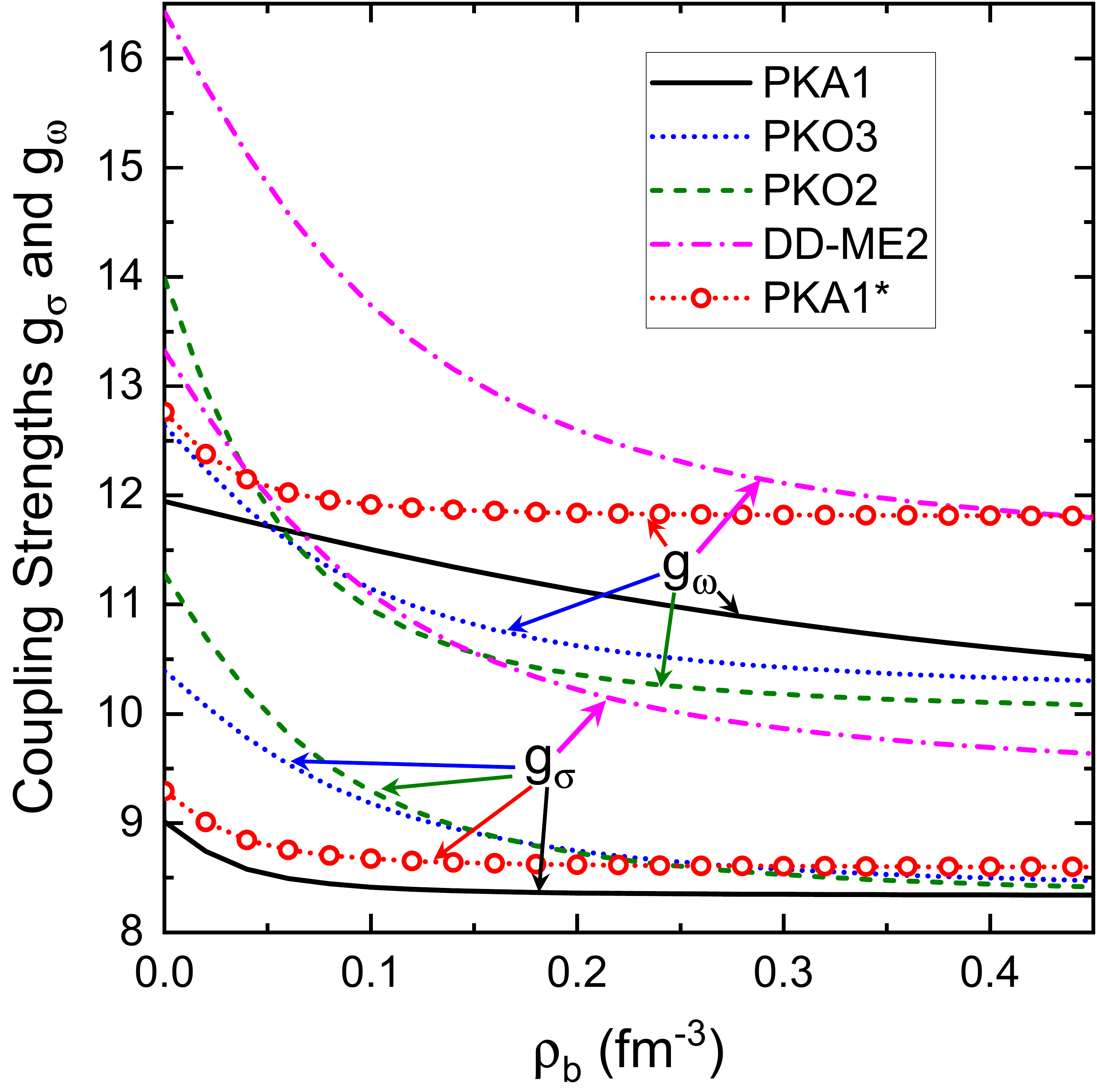}
 \caption{(Color Online) Coupling strengths $g_\sigma$ and $g_\omega$ as functions of density $\rho_b$ (fm$^{-3}$) for PKA1, PKO2, PKO3, DD-ME2 and PKA1*.}\label{fig:coupling}
\end{figure}

{In order to clarify the determinant mechanism, which could be attributed to the considered meson degrees of freedom or the modeling of nuclear in-medium effects, a temporary Lagrangian is deduced from PKA1, namely PKA1* in  Fig. \ref{fig:coupling} that the coupling strengths $g_\sigma$ and $g_\omega$ are set to share the same density dependence (that of original $g_\sigma$ in PKA1), and their values at saturation density are modified simultaneously by several percent to reproduce the binding energy of spherical $^{32}$Mg given by PKA1. Notice that PKA1* is not fully parameterized, but taken as the bridge between PKA1 and PKO3 to illustrate nuclear in-medium effects in determining the deformed ground state of $^{32}$Mg.} From Fig. \ref{Fig:Ebeta} (a), one can see that PKA1*{, which does not support deformed ground state or even a deformed local minimum for $^{32}$Mg,} shows rather similar results as PKO3 during a large range of deformation, roughly $\beta\in\big(0,0.6\big)$. {Combined with Fig. \ref{fig:coupling}, it is then illustrated that the density dependencies of $g_\sigma$ and $g_\omega$, which describe the nuclear in-medium effects carried by the dominant $\sigma$-S and $\omega$-V couplings, are essential for describing correctly the deformed ground state of $^{32}$Mg}, similar as the PSS restoration of high-$l$ PS partners \cite{Geng2019PRC100.051301R} and the liquid-gas critical parameters of thermal nuclear matter \cite{Yang2021PRC103.014304(2021)}.

%As aforementioned, one of the most evident model differences is that PKA1 gives unparalleled density dependent behaviors for the coupling strengths $g_\sigma$ and $g_\omega$, which keep near parallel for the other selected models as shown in Fig. \ref{fig:coupling}. {Combined with the results in Fig. \ref{Fig:Ebeta}, it} enlightened us to deduce a temporary Lagrangian from PKA1, denoted as PKA1* in  Fig. \ref{fig:coupling} that $g_\sigma$ and $g_\omega$ are set to share the same density dependence (that of original $g_\sigma$ in PKA1), and their values at saturation density are modified simultaneously by several percent to reproduce the binding energy of spherical $^{32}$Mg given by original PKA1.

\subsection{ Microscopic evidence for deformed ground state of $^{32}$Mg}

\begin{figure}[h!]\centering
\includegraphics[width=0.92\linewidth]{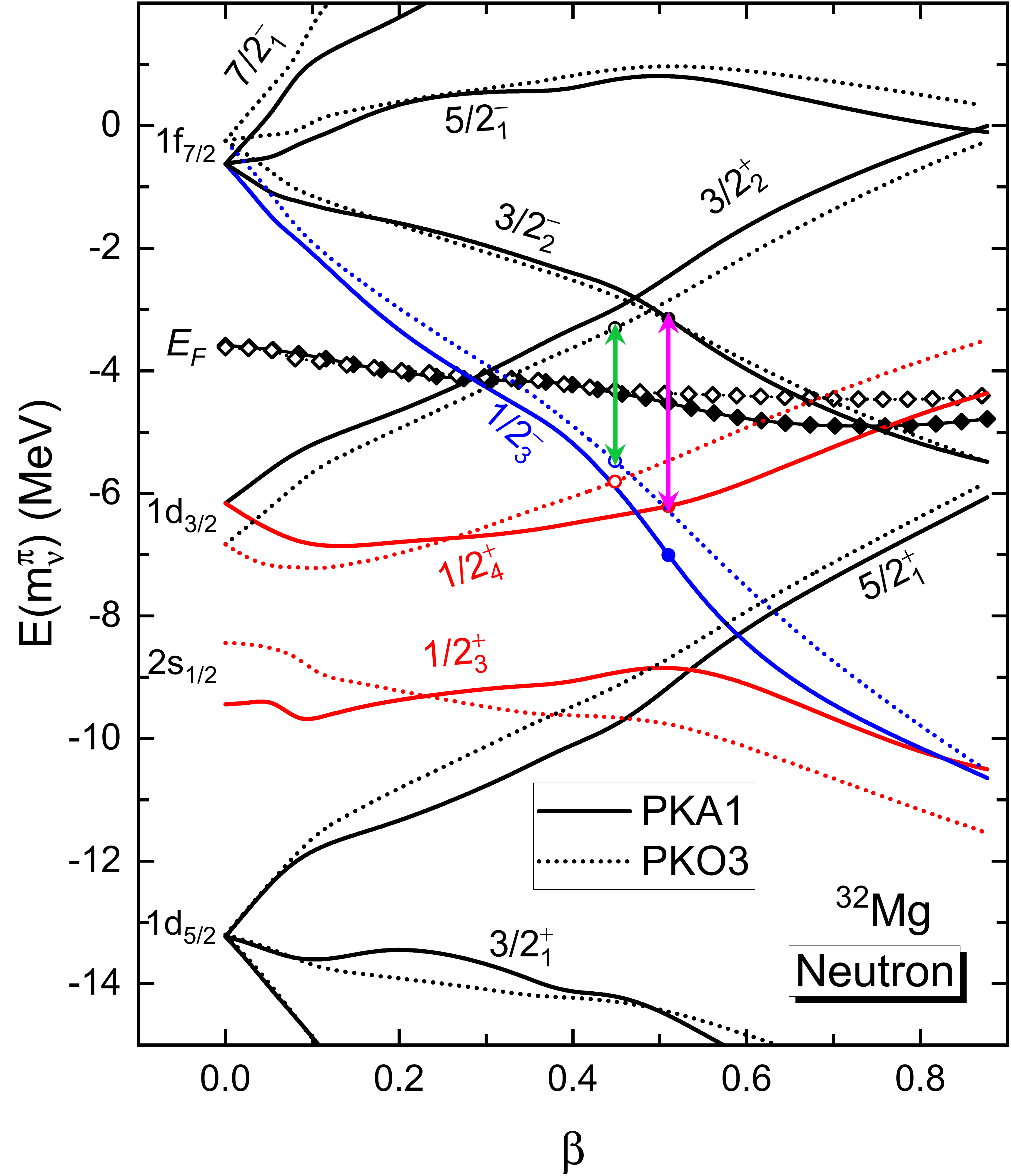}
\caption{(Color Online) Neutron canonical s.p. energies given by PKA1 (solid lines) and PKO3 (dotted lines)
for $^{32}$Mg with respect to the deformation $\beta$, in which the black filled and open diamonds denote the Fermi
energies $E_F$ given by PKA1 and PKO3, respectively. As the references, the positions of prolate minima, reading as $\beta=0.51$ for PKA1 and $0.45$ for PKO3, are marked with filled and open circles, and the arrows for energy gaps there.}\label{Fig:Lev-N}
\end{figure}

{As pointed out in previous subsection, much more enhanced s.p. energy terms $E_{\text{s.p.}}$ given by PKA1 are crucial to give the deformed ground state, see Fig. \ref{Fig:Ebeta} (b). Figure \ref{Fig:Lev-N} further} presents neutron canonical s.p. energies as functions of the deformation $\beta$ for $^{32}$Mg, in which the solid and dashed lines represents the PKA1 and PKO3 results, respectively. For the other selected models including PKA1*, the detailed results are not shown due to their similar systematics as PKO3. In fact, the proton s.p. spectra given by all the selected models show also similar tendency. {It is found in} Fig. \ref{Fig:Lev-N} {that the neutron orbits $1/2_3^+$ and $1/2_4^+$ given by PKA1 and PKO3 are notably different}. For PKO3, on behalf of the other models except PKA1, the orbits $1/2_3^+$ and $1/2_4^+$ show almost monotonous shape evolutions, either decreasing or increasing continuously with respect to the deformation $\beta$. Remind that such behaviors are similar as the ordinary shape evolutions of the s.p. orbits given by Nilsson model \cite{Ring1980Springer-Verlag}.

{However, the shape evolutions of the orbits $1/2_3^+$ and $1/2_4^+$ given by PKA1} are nearly paralleled with each another before reaching the prolate ground state, roughly at $\beta\in(0.1, 0.5)$. Around the Fermi levels $E_F$, due to deeper bound orbit $1/2_4^+$, more remarkable energy gap (marked with arrows) at prolate minimum is obtained by PKA1 with $\beta\approxeq0.51$ (filled circles) than PKO3 with $\beta\approxeq0.45$ (open circles). Moreover, compared to the spherical one $1d_{3/2}$, the valence orbits $1/2_4^+$ and $1/2_3^-$ at prolate minima {are} deeper bound for PKA1{,} but less bound for PKO3. In fact, {this provides a} microscopic evidence for {the emergence of} stably deformed ground state of $^{32}$Mg, in which {the orbit $1/2_4^+$ seems more significant than the odd-parity one $1/2_3^-$}. {Although PKA1 provides more remarkable shell effects than PKO3, the energy gap at the Fermi level is still largely reduced from spherical minimum to prolate ground state, being consistent with the experimental indication that the conventional shell $N=20$ vanishes in $^{32}$Mg.}

\begin{figure}[h!]\centering
\includegraphics[width=0.92\linewidth]{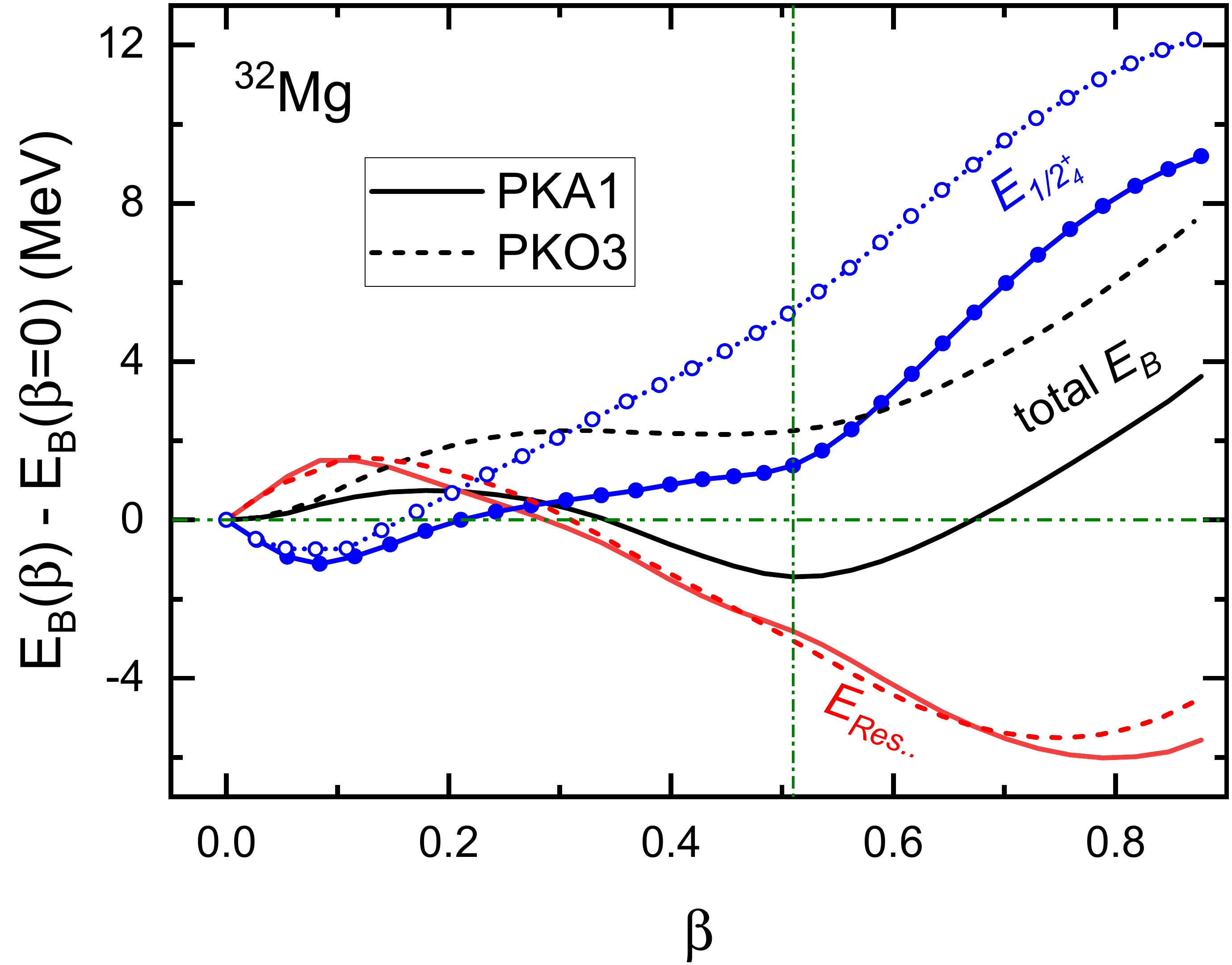}
\caption{(Color Online) Binding energies $E_B$ (MeV) of $^{32}$Mg, the contributions $E_{1/2_4^+}$ from valence orbit $1/2_4^+$ and the rest ones $E_{\text{Res.}}$ with respect to the deformation $\beta$ calculated by PKA1 and PKO3, in which the values at $\beta$ = 0 are taken as the references. See the text for details.}\label{Fig:Cobeta}
\end{figure}

In order to clarify the {role} of the valence orbit $1/2_4^+$, according to Eq. (\ref{eq:sp-energy}), {its} contributions to the binding energy, {denoted as $E_{1/2_4^+}$,} are shown in Fig. \ref{Fig:Cobeta}, in which the solid {lines} (filled circles) and dotted {ones} (open circles) {represent} the PKA1 and PKO3 results, respectively. Except for the $E_{1/2_4^+}$ term, the $E_{\text{Re.}}$, $E_{\text{Oth.}}$ and rest $E_{\text{s.p.}}$ terms in Eq. (\ref{eq:energy1}) are integrated as the term $E_{\text{Res.}} = E_B - E_{1/2_4^+}$ in Fig. \ref{Fig:Cobeta}. Referring to the values at $\beta=0$, it is interesting to see that PKA1 and PKO3 {shows similar evolutions for} the $E_{\text{Res.}}$ terms, but rather different $E_{1/2_4^+}$ ones{, which approximately account} for the {deviations} between PKA1 and PKO3 on the total $E_B$. {Combined with s.p. evolution in Fig. \ref{Fig:Lev-N}, it is clear for the crucial role played by the neutron orbit $1/2_4^+$}.

In the D-RHFB model, the spherical DWS base \cite{Zhou2003PRC68.034323} is used to expand the deformed quasi-particle spinors and canonical s.p. orbits \cite{Geng2020PRC101.064302,Geng2022PRC105.034329}. To better understand the shape evolutions in Fig. \ref{Fig:Lev-N}, Table \ref{tab:Q2} shows the quadruple moment $Q_2$ (fm$^{-2}$) and the proportions (\%) of main spherical components of neutron orbits $1/2_{3}^+$ and $1/2_4^+$ in $^{32}$Mg at the ground state deformation $\beta\approxeq0.51$. {It is seen that} the results given by PKA1 are evidently different from the other selected models which give similar quadruple moments and expansion proportions for both orbits $1/2_3^+$ and $1/2_4^+$. To understand clearly the relations between the $Q_2$ values and expansion proportions, the signs of the couplings between spherical components, whose sum gives the quadruple moment, are list in Table \ref{tab:sign-Q2} as deduced from density expressions in Ref. \cite{Geng2020PRC101.064302} for the orbits $1/2_3^+$ and $1/2_4^+$.

\begin{table}[h]
\caption{Quadruple moment $Q_2$ (fm$^{-2}$) and proportions (in percentage) of the main expansion components of  neutron orbits $1/2_3^+$ and $1/2_4^+$ at $\beta\approx0.51$ for $^{32}$Mg calculated by PKA1, PKO3, PKO2, DD-ME2 and PKA1*.} \label{tab:Q2}\centering
\renewcommand{\arraystretch}{1.25}\setlength{\tabcolsep}{0.4em}
\begin{tabular}{c|c|ccccc}\hline\hline
\multicolumn{2}{c|}{$\beta\approx0.51$}
                                      & PKA1             & PKO3     & PKO2    &  DD-ME2  & PKA1$^*$ \\ \hline
\multirow{4}{*}{$1/2_3^+$} & $Q_2$    & \textbf{  3.46}  &    6.84  &   6.85  &    6.64  &     7.38 \\ \cline{2-7}
                           &$1d_{5/2}$& \textbf{39.0\%}  &  34.0\%  & 30.4\%  &  33.4\%  &   35.4\% \\
                           &$1d_{3/2}$& \textbf{23.5\%}  &  49.7\%  & 49.7\%  &  46.5\%  &   47.0\% \\
                           &$2s_{1/2}$& \textbf{33.9\%}  &  13.7\%  & 17.1\%  &  17.1\%  &   12.2\% \\ \hline\hline
\multirow{4}{*}{$1/2_4^+$} & $Q_2$    &\textbf{$-$2.49}  & $-$5.34  &$-$4.84  & $-$5.13  &  $-$6.34 \\ \cline{2-7}
                           &$1d_{5/2}$& \textbf{ 0.1\%}  &   4.1\%  &  3.7\%  &   3.1\%  &    3.9\% \\
                           &$1d_{3/2}$& \textbf{59.1\%}  &  32.5\%  & 34.9\%  &  36.7\%  &   31.1\% \\
                           &$2s_{1/2}$& \textbf{35.1\%}  &  57.9\%  & 55.6\%  &  53.5\%  &   61.2\% \\ \hline\hline
\end{tabular}
\end{table}

From Table \ref{tab:Q2}, it can be seen that the $Q_2$ values given by PKA1 are much (negatively) smaller than those obtained by the other selected models, which indicates much less deformed $1/2_3^+$ and $1/2_4^+$ orbits for PKA1. In fact, from spherical minimum to prolate ground state, the quadruple moments $Q_2$ of these two orbits given by PKA1 keep rather small values comparing with the other models, i.e., preserving near spherical shapes. {Regarding} the shape consistences with the whole nucleus, it becomes transparent that the s.p. energies of both neutron orbits $1/2_3^+$ and $1/2_4^+$ are slightly and parallelly changed in a fairly wide range of deformation $\beta\in(0.1,0.5)$, see the PKA1 results in Fig. \ref{Fig:Lev-N}. In contrast, with respect to the deformation $\beta$, the models except PKA1 give gradually enlarged prolate and oblate deformations for the orbits $1/2_3^+$ and $1/2_4^+$, respectively, leading to continuously {deep} bound orbit $1/2_3^+$ and high-lying one $1/2_4^+$ as the PKO3 results in Fig. \ref{Fig:Lev-N}.

\begin{table}[h!]
\caption{Signs of the couplings between main spherical components $2s_{1/2}$, $1d_{3/2}$ and $1d_{5/2}$ in quadruple moment $Q_2$ for neutron orbits $1/2_{3}^+$ and $1/2_{4}^+$.}\label{tab:sign-Q2}\centering \renewcommand{\arraystretch}{1.25} \setlength{\tabcolsep}{1em}
\begin{tabular}{c|ccc} \hline\hline
$m^\pi=1/2^+$ & $2s_{1/2}$ & $1d_{3/2}$ & $1d_{5/2}$ \\ \hline
$2s_{1/2}$    &   $0$      &  $-$       &    $+$     \\
$1d_{3/2}$    &   $-$      &  $+$       &    $-$     \\
$1d_{5/2}$    &   $+$      &  $-$       &    $+$     \\ \hline\hline
\end{tabular}
\end{table}

As shown in Table \ref{tab:Q2}, the $1d_{5/2}$ proportions are negligibly small for the orbit $1/2_4^+$, whose $Q_2$ value is therefore decided by mutually cancelled $2s_{1/2}$-$1d_{3/2}$ and $1d_{3/2}$-$1d_{3/2}$ terms given in Table \ref{tab:sign-Q2}. Compared to the other models, negatively reduced $Q_2$ value obtained by PKA1 for the orbit $1/2_4^+$ can be attributed to more cancellation between the $2s_{1/2}$-$1d_{3/2}$ and $1d_{3/2}$-$1d_{3/2}$ terms. Specifically from the other models to PKA1, the increasing $1d_{3/2}$ proportion enlarges the positive $Q_2$ contribution {from} the $1d_{3/2}$-$1d_{3/2}$ {term, and approximately exchanged} $2s_{1/2}$ and $1d_{3/2}$ proportions {lead to roughly unchanged} negative $Q_2$ contribution {from} the $2s_{1/2}$-$1d_{3/2}$ term. Besides, {as seen from Table \ref{tab:Q2}, the orbits $1/2_3^+$ and $1/2_4^+$ show opposite} variation trends from PKA1 to the other models {for} both quadruple moments and expansion proportions. Similarly the $Q_2$ values of the orbit $1/2_3^+$ can be interpreted as well, although the situation looks more complicated than the orbit $1/2_4^+$.

\begin{figure}[h!]\centering
\includegraphics[width=0.92\linewidth]{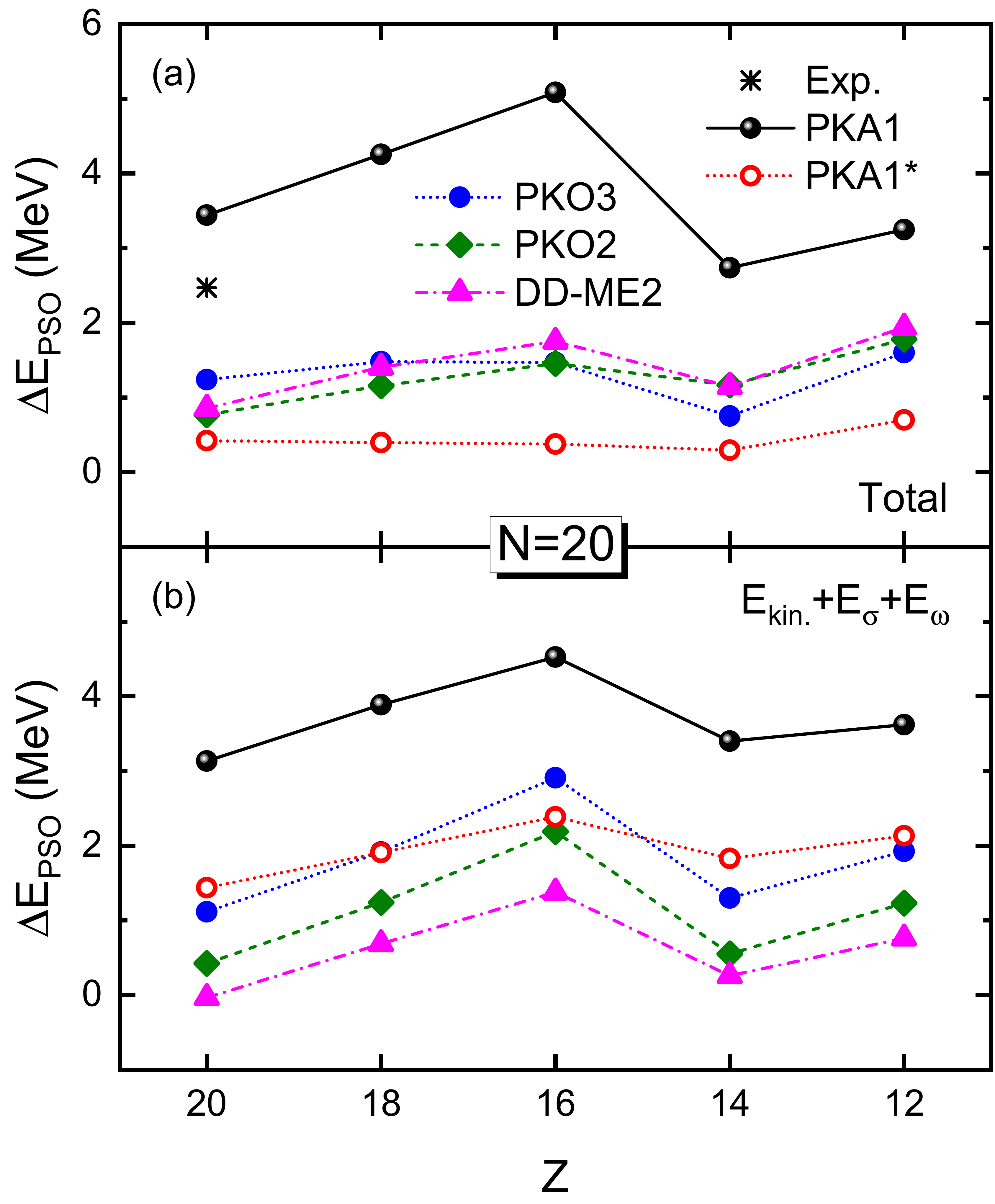}
\caption{(Color Online) Pseudo-spin splitting $\Delta E_{\text{PSO}}$ (MeV) between neutron orbits $1d_{3/2}$ and $2s_{1/2}$ for the $N=20$ isotones with spherical symmetry, including the total [plot (a)] and sum contributions [plot (b)] from kenetic energy, $\sigma$-S and $\omega$-V couplings. The results are given by PKA1, PKO3, PKO2, DD-ME2 and PKA1*. Experimental data is taken form \cite{Grawe2007RPP70.1525}. }\label{Fig:Epso}
\end{figure}

\begin{figure}[h!]\centering
\includegraphics[width=0.92\linewidth]{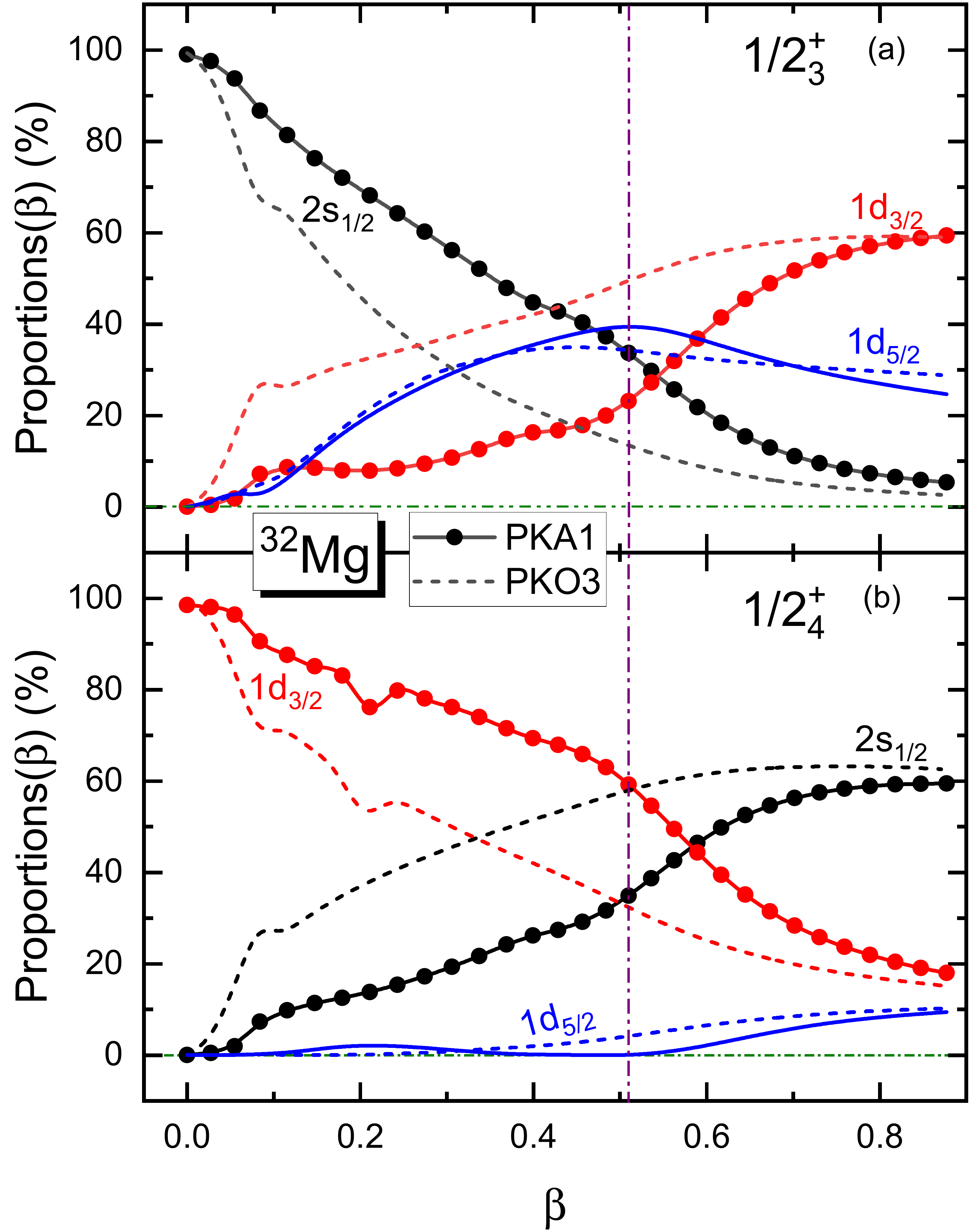}
\caption{(Color Online) Proportions (in percentage) of the main expansion components of  neutron orbits $1/2_3^+$(a) and $1/2_4^+$(b) as functions of deformation parameter $\beta$ for $^{32}$Mg calculated by PKA1 and PKO3.}\label{Fig:xc}
\end{figure}

Therefore, from Tables \ref{tab:Q2} and \ref{tab:sign-Q2}, different {shape} evolution behaviors {given by PKA1 and the other models for} the orbit $1/2_4^+$ {can} be interpreted by {the orbital $Q_2$ values{, which were further explained} by the expansion proportions of} spherical components. %{For instance, compared to the other selected models, PKA1 presents smaller $2s_{1/2}$ and larger $1d_{3/2}$ proportions, leading to less oblately deformed orbit $1/2_4^+$, from which the shape evolution of the orbit $1/2_4^+$ can be understood well}.
{In fact, one may find from Fig. \ref{Fig:Lev-N} that the orbits $1/2_3^+$ and $1/2_4^+$ branch respectively from the spherical PS partners $2s_{1/2}$ and $1d_{3/2}$, and PKA1 presents larger energy gap between these two partners than PKO3. In general, it may affect essentially the mixing of spherical components in both deformed orbits $1/2_3^+$ and $1/2_4^+$. To provide evident proof, applying the selected models, Fig. \ref{Fig:Epso} (a) shows the pseudo-spin orbital (PSO) splittings $\Delta E_{\text{PSO}} =E_{1d_{3/2}} -E_{2s_{1/2}}$ for the $N=20$ isotones from $^{40}$Ca to $^{32}$Mg which are imposed with spherical symmetry. Meanwhile, the dominant contributions, the sum ones from the kinetic energy, $\sigma$-S and $\omega$-V couplings, are shown in Fig. \ref{Fig:Epso} (b). It can be seen in Fig. \ref{Fig:Epso} (a) that PKA1 presents obvious PSS breaking with fairly large $\Delta E_{\text{PSO}}$ values, in contrast to the other models.} %Coincidentally, similar situation has been already found in Table \ref{tab:Q2}, Figs. \ref{Fig:Ebeta} (a) and \ref{fig:coupling}.}

{As supplemental details, Fig. \ref{Fig:xc} shows the proportions of spherical components referring to the deformation $\beta$ for the orbits $1/2_3^+$ [plot (a)] and $1/2_4^+$ [plot (b)]. Being consistent with much enlarged $\Delta E_{\text{PSO}}$ from the other models to PKA1, it is clearly shown in Fig. \ref{Fig:xc}} that PKA1 gives reduced {and retained $1d_{3/2}$-proportion respectively for low-lying orbit $1/2_3^+$ orbit and high-lying one $1/2_4^+$}, and vice versa for the $2s_{1/2}$ proportions. {However}, due to much smaller PSO splittings given by the other models than PKA1 in spherical $^{32}$Mg, the deformation {leads to much more enhanced mixture of the $1d_{3/2}$ component into the low-lying orbit $1/2_3^+$, as well as much larger mixing of the $2s_{1/2}$ ones in the high-lying orbit $1/2_4^+$}, {see both} Table \ref{tab:Q2} and {Fig. \ref{Fig:xc}}. It is worth noting that the PSS breaking described by PKA1 is consistent with the experimental data of $^{40}$Ca \cite{Grawe2007RPP70.1525}, while the other selected models present much reduced $\Delta E_{\text{PSO}}$ values to restore the PSS. {Moreover}, among selected models, only PKA1 can properly reproduce the proton PSO splittings in nearby nuclei $^{40,48}$Ca, see Fig. 1 in Ref. \cite{Li2016PLB753.97}. %As well recognized shell effect is an important mechanism against \delete{the deformation that induces} the mixture of spherical orbital components in deformed orbits.
{Eventually, systematic comparisons between PKA1 and the other selected models prove the fact that the obvious PSS breaking is the microscopic mechanism behind the deformed ground state for $^{32}$Mg.  }

{Moreover, one may notice in} Fig. \ref{Fig:Epso} (b) {that PKA1 and the other selected models present notably different dominant contributions $E_{\text{kin.}} + E_\sigma + E_\omega$ to the PSO splittings $\Delta E_{\text{PSO}}$, which embody the in-medium balance between nuclear attractions and repulsions}. Such in-medium balance, {dominated by} the density dependencies of $g_\sigma$ and $g_\omega$ in Fig. \ref{fig:coupling}, {has been proved to be} a decisive factor for {the PSO splittings} \cite{Geng2019PRC100.051301R}, which measure the PSS breaking. {It is worthwhile to recall the fact in Fig. \ref{Fig:Ebeta} that} the deformed ground state of $^{32}$Mg is not supported by PKA1*, for which the density dependencies of $g_\sigma$ and $g_\omega$ are set as the same. Consistently {as shown in Fig. \ref{Fig:Epso}}, the $\Delta E_{\text{PSO}}$ values are much reduced from PKA1 to PKA1*. Not only for PKA1*, {similar} trends are also seen in Fig. \ref{Fig:Epso} from PKA1 to the models which {fail to reproduce} the deformed ground state of $^{32}$Mg. Regarding similar description as PKO2, PKO3 and DD-ME2, {it is then deduced} that {PKA1* plays the role of a bridge} in verifying the mechanism behind the deformed ground state of $^{32}$Mg, although PKA1* was not fully parameterized.

%In the aforementioned discussions, it has been verified for the tight connections from the proportions of spherical components to the quadruple moments $Q_2$ of the oribts $1/2_3^+$ and $1/2_4^+$ in Table \ref{tab:Q2}, and further to the shape evolutions in Figs. \ref{Fig:Lev-N}-\ref{Fig:Cobeta}. Moreover, the expansion proportions of the orbits are essentially determined by the PSO splittings \insert{shown} in Fig. \ref{Fig:Epso} (a) \insert{and Fig. \ref{Fig:xc}.} Thus, further combined with the results in Fig. \ref{Fig:Ebeta}, it can be concluded that obvious PSS breaking, consistent with the experimental data in nearby isotone $^{40}$Ca, determines the shape evolutions of the s.p. orbits $1/2_3^+$ and $1/2_4^+$ in $^{32}$Mg, being crucial for describing correctly the deformed ground state. Through the entire procedure, PKA1* was taken as a bridge to illustrate a fact that both PSS breaking and deformed ground state of $^{32}$Mg are essentially related with a characteristic in-medium balance between nuclear attractions and repulsions, showing unparalleled density dependent behaviors for the coupling strengths $g_\sigma$ and $g_\omega$.

\section{Summary}\label{sec:Summary}
{Applying the axially deformed relativistic Hartree-Fock-Bogoliubov (D-RHFB) model}, the ground state (GS) of $^{32}$Mg is {studied by performing systematic comparisons between RHF Lagrangian PKA1 and the other selected models, including} the RHF Lagrangians PKO2 and PKO3, and the RMF one DD-ME2. {Restricted with the mean field approach,} only PKA1 presents coincident GS deformation with the experimental measurements for $^{32}$Mg{, in which the modeling of nuclear in-medium effects is found to play a key role}.

{Systematic analysis on neutron single particle structure of $^{32}$Mg show that the valence orbit $1/2_4^+$ with distinctive shape evolution is essential for the formation of deformed GS in $^{32}$Mg. It is illustrated that obvious breaking of the pseudo-spin symmetry (PSS), found in $^{32}$Mg and nearby $N=20$ isotones, plays a decisive role on the mixing of spherical pseudo-spin partners $2s_{1/2}$ and $1d_{3/2}$ in the orbit $1/2_4^+$ and further on its shape evolution. It is thus revealed for obvious PSS breaking as the microscopic mechanism behind the deformed GS of $^{32}$Mg.}

{Consistent with existing experimental data, obvious PSS breaking in $^{32}$Mg and nearby $N=20$ isotones can be deduced from the modeling of nuclear in-medium effects described by PKA1, which is characterized by unparalleled density-dependent behaviors of the coupling strengths $g_\sigma$ and $g_\omega$ in dominant meson-nucleon coupling channels. Perspectively, the revealed mechanism that promises the deformed GS of $^{32}$Mg, particularly the embedded nuclear in-medium effects, provides qualitative guidance on understanding the nature of nuclear force.}

\begin{acknowledgements}
%If you'd like to thank anyone, place your comments here
%and remove the percent signs.
This work is partly supported by the National Natural Science Foundation of China under Grant Nos. 12275111 and 12075104, the Strategic Priority Research Program of Chinese Academy of Sciences under Grant No. XDB34000000, the National Key Research and Development (R\&D) Program under Grant No. 2021YFA1601500, and the Supercomputing Center of Lanzhou University.

\end{acknowledgements}

% BibTeX users please use one of
%\bibliographystyle{spbasic}      % basic style, author-year citations
%\bibliographystyle{spmpsci}      % mathematics and physical sciences
%\bibliographystyle{spphys}       % APS-like style for physics
%\bibliography{RHFB}   % name your BibTeX data base
%\end{document}

% Non-BibTeX users please use

\end{document}